\documentclass[%
aip,
floatfix,
preprintnumbers,
reprint,
superscriptaddress,
amsmath, amssymb,
longbibliography,
]{revtex4-1}

\usepackage{soul}

\usepackage[utf8]{inputenc}
\usepackage{graphicx}
\usepackage{bm}
\usepackage[breaklinks=true]{hyperref}
\hypersetup{bookmarksnumbered, pdfpagemode=UseOutlines, 
	pdfauthor={K.\ S.\ Das}, 
	pdftitle={Temperature dependence of the effective spin-mixing conductance probed with lateral non-local spin valves}, 
	pdfdisplaydoctitle, 
	colorlinks=true, citecolor=blue, filecolor=blue, linkcolor=blue, urlcolor=blue}

\begin{document}
	
	\title{Temperature dependence of the effective spin-mixing conductance probed with lateral non-local spin valves}
	
	\author{K.\ S.\ \surname{Das}}
	\email[e-mail: ]{K.S.Das@rug.nl}
	\affiliation{Physics of Nanodevices, Zernike Institute for Advanced Materials, University of Groningen, 9747 AG Groningen, The Netherlands}
	\author{F.\ K.\ Dejene}
	\affiliation{Department of Physics, Loughborough University, Loughborough LE11 3TU, United Kingdom}
	\author{B.\ J.\ \surname{van Wees}}
	\affiliation{Physics of Nanodevices, Zernike Institute for Advanced Materials, University of Groningen, 9747 AG Groningen, The Netherlands}
	\author{I.\ J.\ Vera-Marun}
	\email[e-mail: ]{ivan.veramarun@manchester.ac.uk}
	\affiliation{School of Physics and Astronomy, University of Manchester, Manchester M13 9PL, United Kingdom}
	
	
	\begin{abstract}
		We report the temperature dependence of the effective spin-mixing conductance between a normal metal (aluminium, Al) and a magnetic insulator ($\text{Y}_3\text{Fe}_5\text{O}_{12}$, YIG). Non-local spin valve devices, using Al as the spin transport channel, were fabricated on top of YIG and SiO$_2$ substrates. By comparing the spin relaxation lengths in the Al channel on the two different substrates, we calculate the effective spin-mixing conductance ($G_\text{s}$) to be $3.3\times10^{12}$~$\Omega^{-1}\text{m}^{-2}$ at 293~K for the Al/YIG interface. A decrease of up to 84\% in $G_\text{s}$ is observed when the temperature ($T$) is decreased from 293~K to 4.2~K, with $G_\text{s}$ scaling with $(T/T_\text{c})^{3/2}$. The real part of the spin-mixing conductance ($G_\text{r}\approx 5.7\times10^{13}~ \Omega^{-1}\text{m}^{-2}$), calculated from the experimentally obtained $G_\text{s}$, is found to be approximately independent of the temperature. We evidence a hitherto unrecognized underestimation of $G_\text{r}$ extracted from the modulation of the spin signal by rotating the magnetization direction of YIG with respect to the spin accumulation direction in the Al channel, which is found to be 50 times smaller than the calculated value.     
	\end{abstract}
	
	
	
	\maketitle
	
	
	The transfer of spin information between a normal metal (NM) and a magnetic insulator (MI) is the crux of electrical injection and detection of spins in the rapidly emerging fields of magnon spintronics \cite{chumak_magnon_2015} and antiferromagnetic spintronics \cite{jungwirth_antiferromagnetic_2016,baltz_antiferromagnetic_2018}. The spin current flowing through the NM/MI interface is governed by the spin-mixing conductance \cite{brataas_finite-element_2000,brataas_non-collinear_2006,takahashi_spin_2010,jia_spin_2011}, $G_{\uparrow\downarrow}$, which plays a crucial role in spin transfer torque \cite{stiles_anatomy_2002,ralph_spin_2008,brataas_current-induced_2012}, spin pumping \cite{tserkovnyak_enhanced_2002,deorani_role_2013}, spin Hall magnetoresistance (SMR) \cite{nakayama_spin_2013,vlietstra_spin-hall_2013} and spin Seebeck experiments \cite{uchida_spin_2010}. In these experiments, the spin-mixing conductance $(G_{\uparrow\downarrow}=G_\text{r}+iG_\text{i})$, composed of a real ($G_\text{r}$) and an imaginary part ($G_\text{i}$), determines the transfer of spin angular momentum between the spin accumulation ($\vec{\mu}_\text{s}$) in the NM and the magnetization ($\vec{M}$) of the MI in the \textit{non-collinear} case. However, recent experiments on the spin Peltier effect \cite{flipse_observation_2014}, spin sinking \cite{dejene_control_2015} and non-local magnon transport in magnetic insulators \cite{cornelissen_long-distance_2015,cornelissen_magnon_2016} necessitate the transfer of spin angular momentum through the NM/MI interface also in the \textit{collinear} case ($\vec{\mu}_\text{s} \parallel \vec{M}$). This is taken into account by the effective spin-mixing conductance ($G_\text{s}$) concept, according to which the transfer of spin angular momentum across the NM/MI interface can occur, irrespective of the mutual orientation between $\vec{\mu}_\text{s}$ and $\vec{M}$, via local thermal fluctuations of the equilibrium magnetization (thermal magnons \cite{kalinikos_theory_1986}) in the MI. The spin current density ($\vec{j}_\text{s}$) through the NM/MI interface can, therefore, be expressed as \cite{dejene_control_2015,brataas_spin_2012,chen_theory_2013}:
	\begin{equation}
	 \vec{j}_\text{s}=G_\text{r}\hat{m}\times\left(\vec{\mu}_\text{s}\times\hat{m}\right)+G_\text{i}\left(\vec{\mu}_\text{s}\times\hat{m}\right)+G_\text{s}\vec{\mu}_\text{s},
	 \label{eq:SpinCurrentEq}
	 \end{equation}
	 where, $\hat{m}$ is a unit vector pointing along the direction of $\vec{M}$. While $G_\text{r}$ and $G_\text{i}$ have been extensively studied in spin torque and SMR experiments \cite{xia_spin_2002,vlietstra_exchange_2013,meyer_temperature_2014}, direct experimental studies on the temperature dependence of $G_\text{s}$ are lacking.  
	
	In this letter, we report the first systematic study of $G_\text{s}$ versus temperature ($T$) for a NM/MI interface. For this, we utilize the lateral non-local spin valve (NLSV) geometry, which provides an alternative way to study the spin-mixing conductance using pure spin currents in a NM with low spin-orbit coupling (SOC) \cite{villamor_modulation_2015,dejene_control_2015,muduli_detection_2018}. A low SOC of the NM in the NLSV technique also ensures that the spin-mixing conductance is not overestimated due to spurious proximity effects in NMs with high SOC or close to the Stoner criterion, such as Pt \cite{huang_transport_2012,lu_hybrid_2013,cahaya_spin_2014}. We exclusively address the temperature dependence of $G_\text{s}$ for the aluminium (Al)/$\text{Y}_3\text{Fe}_5\text{O}_{12}$ (YIG) interface, which is obtained by comparing the spin relaxation length ($\lambda_\text{N}$) in similar Al channels on a magnetic YIG substrate and a non-magnetic SiO$_2$ substrate, as a function of temperature. $G_\text{s}$ decreases by about $84\%$ when the temperature is decreased from 293~K to 4.2~K and scales with $(T/T_\text{c})^{3/2}$, where $T_\text{c}=560$~K is the Curie temperature of YIG, consistent with theoretical predictions \cite{cornelissen_magnon_2016,zhang_spin_2012,bender_interfacial_2015,xiao_transport_2015}. The real part of the spin-mixing conductance ($G_\text{r}$) is then calculated from the experimentally obtained values of $G_\text{s}$ and compared with the modulation of the spin signal in rotation experiments, where the magnetization direction of YIG ($\vec{M}$) is rotated with respect to $\vec{\mu}_\text{s}$.

	\begin{figure*}[tbp]
		\includegraphics*[angle=0, trim=0mm 0mm 0mm 0mm, width=170mm]{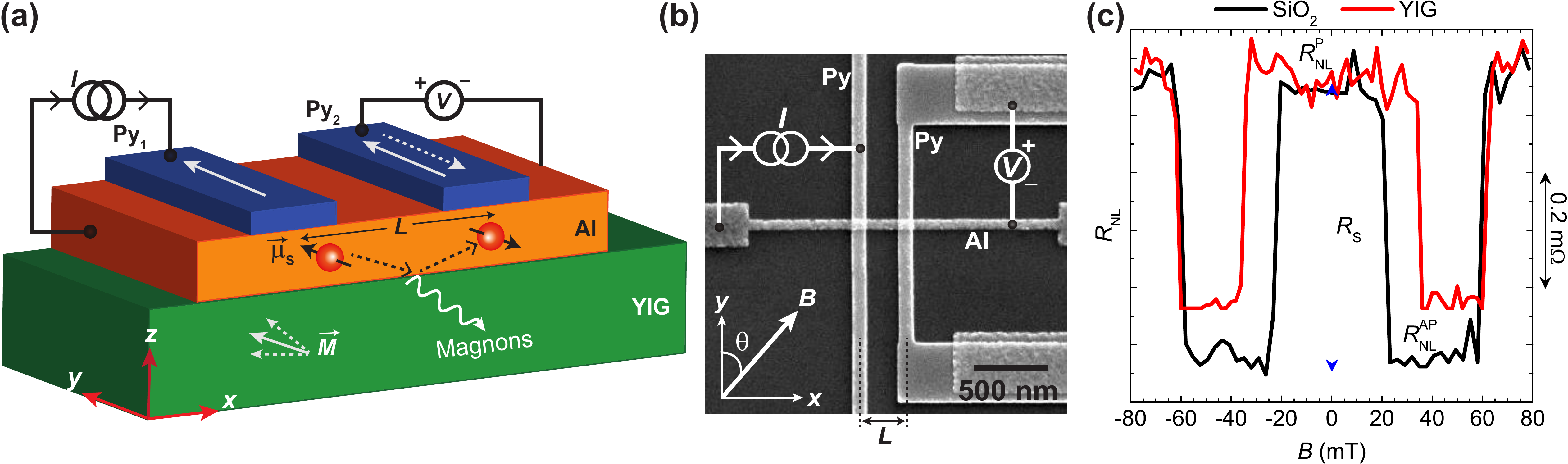}
		\caption{
			\label{fig:DeviceGeometry}
			\textbf{(a)} Schematic illustration of the experimental geometry. The spin accumulation ($\vec{\mu}_\text{s}$), injected into the Al channel by the Py injector, has an additional relaxation pathway into the (insulating) magnetic YIG substrate due to local thermal fluctuations of the equilibrium YIG magnetization ($\vec{M}$) or thermal magnons. \textbf{(b)} SEM image of a representative NLSV device along with the illustration of the electrical connections for the NLSV measurements. An alternating current ($I$) was sourced from the left Py strip (injector) to the left end of the Al channel and the non-local voltage ($V_\text{NL}$) was measured across the right Py strip (detector) with reference to the right end of the Al channel. An external magnetic field ($B$) was swept along the $y$-axis in the non-local spin valve (NLSV) measurements. In the rotation measurements, $B$ was applied at different angles ($\theta$) with respect to the $y$-axis in the $xy$-plane. \textbf{(c)} NLSV measurements at $T=293$~K for an Al channel length ($L$) of 300~nm on the YIG substrate (red) and on the SiO$_2$ substrate (black). 
		}
	\end{figure*}
	
	The NLSVs with Al spin transport channel were fabricated on top of YIG and SiO$_2$ thin films in multiple steps using electron beam lithography (EBL), electron beam evaporation of the metallic layers and resist lift-off technique, following the procedure described in Ref.~\onlinecite{das_anisotropic_2016}. The 210~nm thick YIG film on Gd$_3$Ga$_5$O$_{12}$ substrate and the 300~nm thick SiO$_2$ film on Si substrate were obtained commercially from Matesy GmbH and Silicon Quest International, respectively. Permalloy (Ni$_{80}$Fe$_{20}$, Py) has been used as the ferromagnetic electrodes for injecting and detecting a non-equilibrium spin accumulation in the Al channel. A 3~nm thick Ti underlayer was deposited prior to the evaporation of the 20~nm thick Py electrodes. The Ti underlayer prevents direct injection and detection of spins in the YIG substrate via the anomalous spin Hall effect in Py \cite{das_spin_2017,das_efficient_2018}. \textit{In-situ} $\text{Ar}^+$ ion milling for 20~seconds at an Ar gas pressure of $4\times10^{-5}$~Torr was performed, prior to the evaporation of the 55~nm thick Al channel, ensuring a transparent and clean Py/Al interface. A schematic of the device geometry is depicted in Fig.~\ref{fig:DeviceGeometry}(a) and a scanning electron microscope (SEM) image of a representative device is shown in Fig.~\ref{fig:DeviceGeometry}(b). A low frequency (13~Hz) alternating current source ($I$) with an r.m.s. amplitude of 400~$\mu$A was connected between the left Py strip (injector) and the left end of the Al channel. The non-local voltage ($V_\text{NL}$) due to the non-equilibrium spin accumulation in the Al channel was measured between the right Py strip (detector) and the right end of the Al channel using a standard lock-in technique. The measurements were carried out under a low vacuum atmosphere in a variable temperature insert, placed within a superconducting magnet.
	
	\begin{figure*}[tbp]
		\includegraphics*[angle=0, trim=0mm 0mm 0mm 0mm, width=170mm]{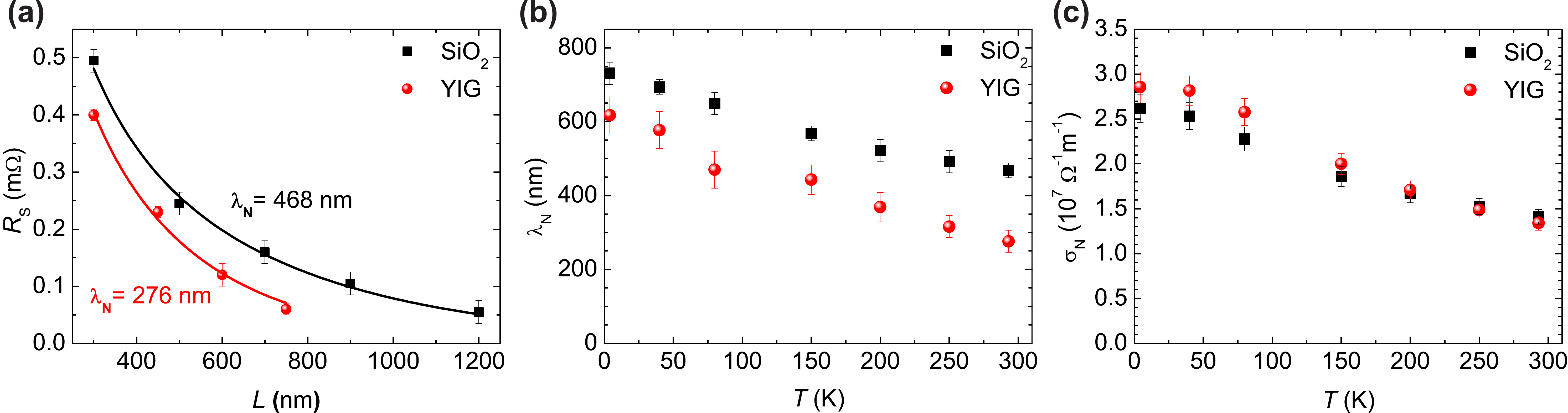}
		\caption{
			\label{fig:SpinValve}
			\textbf{(a)} The spin signal ($R_\text{s}$) plotted as a function of the Al channel length ($L$) for NLSV devices on YIG (red circles) and SiO$_2$ (black square) substrates at 293~K. The solid lines represent the fits to the spin diffusion model (Eq.~\ref{eq:1DspinDiffusion}). \textbf{(b)} The effective spin relaxation length in the Al channel ($\lambda_\text{N}$) extracted at different temperatures ($T$). $\lambda_\text{N}$ is smaller on the YIG substrate as compared to the SiO$_2$ substrate. \textbf{(c)} The electrical conductivity ($\sigma_\text{N}$) of the Al channels on the YIG and the SiO$_2$ substrates as a function of temperature. The close match between the two conductivities suggests similar quality of the Al film grown on both substrates.
		}
	\end{figure*}
	
	In the NLSV measurements, an external magnetic field ($B$) was swept along the $y$-axis and the corresponding non-local resistance ($R_\text{NL}=V_\text{NL}/I$) was measured. In Fig.~\ref{fig:DeviceGeometry}(c), NLSV measurements for an Al channel length ($L$) of 300~nm at $T=293$~K are shown for two devices, one on YIG (red) and another on SiO$_2$ (black). The spin signal, $R_\text{s}=R_\text{NL}^\text{P}-R_\text{NL}^\text{AP}$, is defined as the difference in the two distinct states corresponding to the parallel ($R_\text{NL}^\text{P}$) and the anti-parallel ($R_\text{NL}^\text{AP}$) alignment of the Py electrodes' magnetizations. The $R_\text{s}$ was measured as a function of the separation ($L$) between the injector and the detector electrodes for several devices fabricated on YIG and SiO$_2$ substrates, as shown in Fig.~\ref{fig:SpinValve}(a). To  determine the spin relaxation length ($\lambda_\text{N}$) in the Al channels on YIG ($\lambda_\text{N, YIG}$) and SiO$_2$ ($\lambda_{\text{N, }\text{SiO}_2}$) substrates, the experimental data in Fig.~\ref{fig:SpinValve}(a) were fitted with the spin diffusion model \cite{takahashi_spin_2003} for transparent contacts:
	\begin{equation}
	R_\text{s}=\frac{4\alpha_{\text{F}}^2}{(1-\alpha_{\text{F}}^2)^2}{\cal R}_\text{N}\left(\frac{{\cal R}_\text{F}}{{\cal R}_\text{N}}\right)^2\frac{e^{-L/\lambda_\text{N}}}{1-e^{-2L/\lambda_\text{N}}},
	\label{eq:1DspinDiffusion}
	\end{equation}
	where, $\alpha_{\text{F}}$ is the bulk spin polarization of Py, ${\cal R}_\text{N}=\rho_\text{N} \lambda_\text{N} / w_\textrm{N} t_\text{N}$ and ${\cal R}_\text{F}=\rho_\text{F} \lambda_\text{F} / w_\text{N} w_\text{F}$ are the spin resistances of Al and Py, respectively. $\lambda_\text{N(F)}$, $\rho_\text{N(F)}$, $w_\textrm{N(F)}$ and $t_\text{N}$ are the spin relaxation length, electrical resistivity, width and thickness of Al (Py), respectively. At room temperature, $\lambda_\text{N, YIG}=(276\pm30)$~nm and $\lambda_{\text{N, }\text{SiO}_2}=(468\pm20)$~nm were extracted, with $\alpha_{\text{F}}\lambda_\text{F}=(0.84\pm0.05)$~nm. 
	
	The NLSV measurements were carried out at different temperatures, enabling the extraction of $\lambda_\text{N, YIG}$ and $\lambda_{\text{N, }\text{SiO}_2}$, as shown in Fig.~\ref{fig:SpinValve}(b). From this temperature dependence, it is obvious that $\lambda_\text{N, YIG}$ is lower than $\lambda_{\text{N, }\text{SiO}_2}$ throughout the temperature range of 4.2~K to 293~K. The corresponding electrical conductivities of the Al channel ($\sigma_\text{N}$) on the two different substrates were also measured by the four-probe technique as a function of $T$, as shown in Fig.~\ref{fig:SpinValve}(c). The similar values of $\sigma_\text{N}$ for the Al channels on both YIG and the SiO$_2$ substrates suggests that there is no significant difference in the structure and quality of the Al films between the two substrates. Therefore, considering the dominant Elliott-Yafet spin relaxation mechanism in Al \cite{zutic_spintronics:_2004}, differences in the spin relaxation rate within the Al channels cannot account for the difference in the effective spin relaxation lengths between the two substrates. 
	
	The smaller values of $\lambda_\text{N, YIG}$ as compared to $\lambda_{\text{N, }\text{SiO}_2}$ suggest that there is an additional spin relaxation mechanism for the spin accumulation in the Al channel on the magnetic YIG substrate. This is expected via additional spin-flip scattering at the Al/YIG interface, mediated by thermal magnons in YIG and governed by the effective spin-mixing conductance ($G_\text{s}$). As described in Ref.~\onlinecite{dejene_control_2015}, $\lambda_\text{N, YIG}$ and $\lambda_{\text{N, }\text{SiO}_2}$ are related to $G_\text{s}$ as
	\begin{equation}
	\frac{1}{\lambda^2_\text{N, YIG}}=\frac{1}{\lambda^2_{\text{N, }\text{SiO}_2}}+\frac{1}{\lambda^2_\text{r}},
	\label{eq:GsCalculate}
	\end{equation}  
	where, $\lambda_\text{r}=2G_\text{s}/(t_\text{Al}\sigma_\text{N})$. Using the extracted values of $\lambda_\text{N}$ from Fig.~\ref{fig:SpinValve}(b) and the measured values of $\sigma_\text{N}$ for the devices on YIG from Fig.~\ref{fig:SpinValve}(c), we calculate $G_\text{s}=3.3\times10^{12}$~$\Omega^{-1}\text{m}^{-2}$ at 293~K. At 4.2~K, $G_\text{s}$ decreases by about $84\%$ to $5.4\times10^{11}$~$\Omega^{-1}\text{m}^{-2}$. 
	
\begin{figure}[b]
		\includegraphics*[angle=0, trim=0mm 0mm 0mm 0mm, width=85mm]{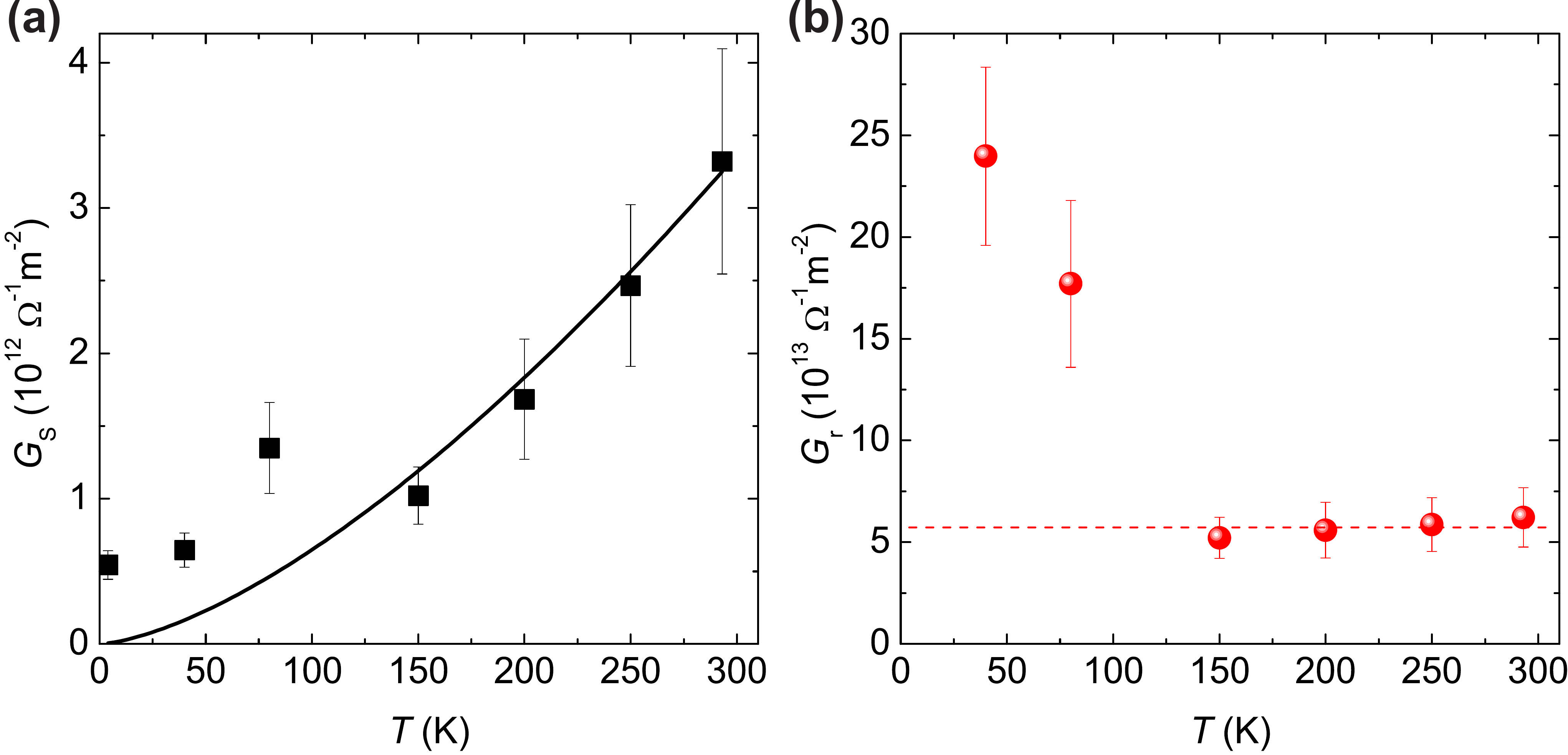}
		\caption{
			\label{fig:GsGr-T}
			\textbf{(a)} Temperature dependence of the effective spin-mixing conductance (black symbols). $G_\text{s}$ scales with the temperature as $(T/T_\text{c})^{3/2}$ (solid line). \textbf{(b)} The real part of the spin-mixing conductance ($G_\text{r}$) is calculated from Eq.~\ref{eq:GsandGr} by using the experimentally obtained values of $G_\text{s}$. $G_\text{r}$ $(\approx 5.7\times10^{13}~ \Omega^{-1}\text{m}^{-2})$ is essentially found to be constant (dashed line) for $T>100~K$. 
		}
	\end{figure}

	The temperature dependence of $G_\text{s}$ is shown in Fig.~\ref{fig:GsGr-T}(a). Since the concept of the effective spin-mixing conductance is based on the thermal fluctuation of the magnetization (thermal magnons), $G_\text{s}$ is expected to scale as $(T/T_\text{c})^{3/2}$, where $T_\text{c}$ is the Curie temperature of the magnetic insulator \cite{takahashi_spin_2010,zhang_spin_2012,bender_interfacial_2015,cornelissen_magnon_2016}. Using $T_\text{c}=560$~K for YIG, we fit the experimental data to $C(T/T_\text{c})^{3/2}$, which is depicted as the solid line in Fig.~\ref{fig:GsGr-T}(a). The temperature independent prefactor, $C$, was found to be $8.6\times10^{12}$~$\Omega^{-1}\text{m}^{-2}$. The agreement with the experimental data confirms the expected scaling of $G_\text{s}$ with temperature. Note that the deviation from the $(T/T_\text{c})^{3/2}$ scaling at lower temperatures could be in part due to slightly different quality of the Al film on the YIG substrate. Nevertheless, the small difference of $\approx10\%$ in the electrical conductivities of the Al channel on the two different substrates at $T<100$~K in Fig.~\ref{fig:SpinValve}(c) cannot account for the differences in $\lambda_\text{N}$. On the other hand, we note that quantum magnetization fluctuations \cite{zholud_spin_2017,kim_manifestation_2014} in YIG can also play a role at low $T$, leading to an enhanced $G_\text{s}$.       
	
	\begin{figure*}[tbp]
		\includegraphics*[angle=0, trim=0mm 0mm 0mm 0mm, width=170mm]{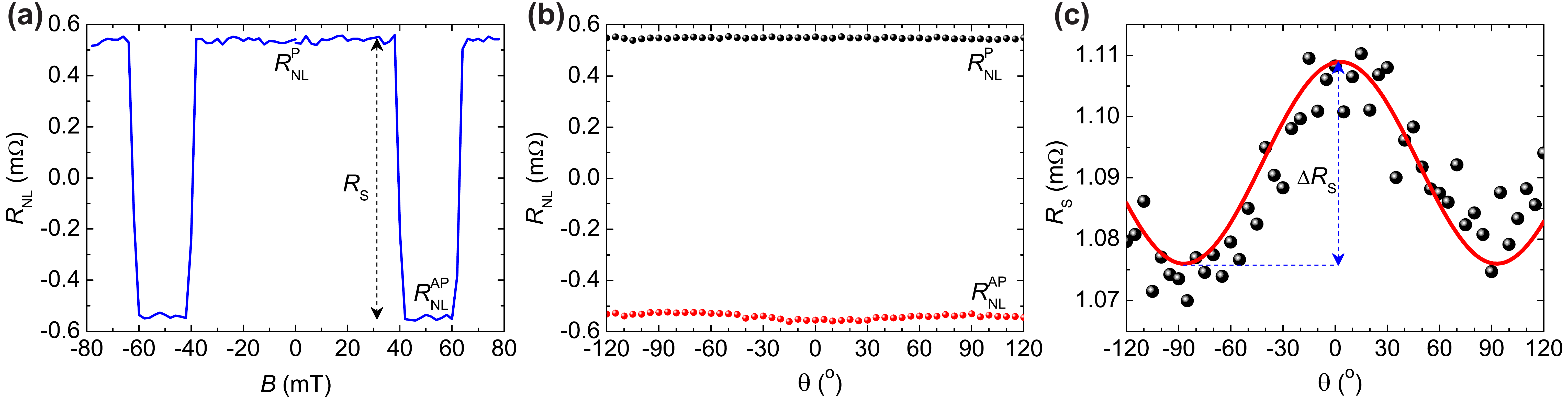}
		\caption{
			\label{fig:GR}
			\textbf{(a)} NLSV measurement for a device on the YIG substrate with $L=300$~nm at 150~K. \textbf{(b)} Rotation measurement for the same device with $B=20$~mT applied at different angles ($\theta$) with respect to the $y$-axis. The black and the red symbols correspond to the average of ten rotation measurements carried out with the magnetization of the Py electrodes in the parallel (P) and the anti-parallel (AP) configurations, respectively. \textbf{(c)} The spin signal ($R_\text{s}=R_\text{NL}^\text{P}-R_\text{NL}^\text{AP}$) exhibits a periodic modulation of magnitude $\Delta R_\text{s}$ when the angle $\theta$ between the magnetization direction in YIG ($\vec{M}$) and the spin accumulation direction in Al ($\vec{\mu}_\text{s}$) is changed. The black symbols represent the experimental data at 150~K, while the red line is the numerical modelling result corresponding to $G_\text{r}=1\times10^{12}$~$\Omega^{-1}\text{m}^{-2}$.    
		}
	\end{figure*}
	
	Next, we investigate the temperature dependence of the real part of the spin-mixing conductance ($G_\text{r}$). For this, we first calculate $G_\text{r}$ from the experimentally obtained $G_\text{s}$, using the following expression \cite{cornelissen_magnon_2016}: 
	\begin{equation}
    G_\text{s}=\frac{3\zeta(3/2)}{2\pi s \Lambda^3}G_\text{r},
    \label{eq:GsandGr}
	\end{equation} 
	where $\zeta(3/2)=2.6124$ is the Riemann zeta function calculated at $3/2$, $s=S/a^3$ is the spin density with total spin $S=10$ in a unit cell of volume $a^3=1.896~\text{nm}^3$, and $\Lambda=\sqrt{4\pi D_\text{s}/k_\text{B}T}$ is the thermal de Broglie wavelength for magnons, with $D_\text{s}=8.458\times10^{-40}~\text{Jm}^2$ being the spin wave stiffness constant for YIG \cite{cherepanov_saga_1993,cornelissen_magnon_2016}. The temperature dependence of the calculated $G_\text{r}$ is shown in Fig.~\ref{fig:GsGr-T}(b). Keeping in mind that Eq.~\ref{eq:GsandGr} is not valid in the limits of $T \to T_\text{c}$ and $T \to 0$, we ignore the data points below 100~K. Above this temperature, $G_\text{r}$ is almost constant at $\approx 5.7\times10^{13}~ \Omega^{-1}\text{m}^{-2}$, represented by the dashed line in Fig.~\ref{fig:GsGr-T}(b). This is consistent with Ref.~\onlinecite{meyer_temperature_2014}, where $G_\text{r}$ was found to be $T$-independent. Moreover, the magnitude of $G_\text{r}$ is comparable with previously reported values for Al/YIG \cite{dejene_control_2015} and Pt/YIG \cite{cornelissen_magnon_2016,velez_competing_2016} interfaces.
	
	An alternative approach for extracting $G_\text{r}$ from the NLSVs fabricated on the YIG substrate, is by the rotation of the sample with respect to a low magnetic field in the $xy$-plane. We have also followed this method, described in Refs.~\onlinecite{dejene_control_2015,villamor_modulation_2015}. In the rotation experiments, the angle $\theta$ between the magnetization direction in YIG ($\vec{M}$) and the spin accumulation direction in Al ($\vec{\mu}_\text{s}$) is changed, which results in the modulation of the spin signal in the Al channel due to the transfer of spin angular momentum across the Al/YIG interface, as described in Eq.~\ref{eq:SpinCurrentEq}, dominated by the $G_\text{r}$ term. First, the NLSV measurement for a device with $L=300$~nm was carried out at 150~K, as shown in Fig.~\ref{fig:GR}(a). In the next step, $B=20$~mT was applied in the $xy$-plane and the sample was rotated, with the magnetization orientations of the Py electrodes set in the parallel (P) or the anti-parallel (AP) configuration. For improving the signal-to-noise ratio, ten measurements were performed for each of the configurations (P and AP). The average of these measurements is shown in Fig.~\ref{fig:GR}(b). The spin signal is extracted from Fig.~\ref{fig:GR}(b) and plotted as a function of $\theta$ in Fig.~\ref{fig:GR}(c). $R_\text{s}$ exhibits a periodic modulation with the maxima at $\theta=0^\circ$ and minima at $\theta=\pm90^\circ$, consistent with the behaviour predicted in Eq.~\ref{eq:SpinCurrentEq}. The modulation in the $R_\text{s}$, defined as $\frac{(R_\text{s}^{0^\circ}-R_\text{s}^{\pm90^\circ})}{R_\text{s}^{0^\circ}}=\frac{\Delta R_\text{s}}{R_\text{s}^{0^\circ}}$, was found to be $2.8\%$. A similar modulation of $2.9\%$ was reported in Ref~\onlinecite{villamor_modulation_2015} for an NLSV with a Cu channel on YIG with $L=570$~nm at the same temperature.
	
	$G_\text{r}$ is estimated from the rotation measurements using 3D finite element modelling, as described in Ref.~\onlinecite{dejene_control_2015}. From the modelled curve for the spin signal modulation, shown as the red line in Fig.~\ref{fig:GR}(c), we extract $G_\text{r}=1\times10^{12}$~$\Omega^{-1}\text{m}^{-2}$. This value is comparable to that reported in Ref.~\onlinecite{villamor_modulation_2015}, within a factor of 2, for an evaporated Cu channel on YIG. However, this value is more than 50 times smaller than our estimated value from Eq.~\ref{eq:GsandGr}, and also that reported in Ref.~\onlinecite{dejene_control_2015} for a sputtered Al channel on YIG. One reason behind the small magnitude of $G_\text{r}$ extracted from the rotation measurements can be attributed to the thin film deposition technique used. In Ref.~\onlinecite{vlietstra_spin-hall_2013}, it was shown that the SMR signal for a sputtered Pt film on YIG is about an order of magnitude larger than that for an evaporated Pt film. Moreover, during the fabrication of our NLSVs, an $\text{Ar}^+$ ion milling step is carried out prior to the evaporation of the NM channel for ensuring a clean interface between the NM and the ferromagnetic injector and detector electrodes \cite{dejene_control_2015,villamor_modulation_2015}. Consequently, this also leads to the milling of the YIG surface on which the NM is deposited, resulting in the formation of an $\approx2$~nm thick amorphous YIG layer at the interface \cite{casanova_private_2018}. Since an external magnetic field of 20~mT is not sufficient to completely align the magnetization direction within this amorphous layer parallel to the field direction \cite{wesenberg_long-distance_2017}, the resulting modulation in the spin signal will be smaller. This might lead to the underestimation of $G_\text{r}$. Note that since the effect of $G_\text{s}$ does not depend on the magnetization orientation of YIG (Eq.~\ref{eq:SpinCurrentEq}), the milling does not affect the estimation of $G_\text{s}$. Our observations are consistent with a similarly small value of $G_\text{r}\approx4\times10^{12}$~$\Omega^{-1}\text{m}^{-2}$ reported in Ref.~\onlinecite{villamor_modulation_2015} for the Cu/YIG interface, where the Cu channel was evaporated following a similar $\text{Ar}^+$ ion milling step. Using the reported values of $\lambda_\text{N}=522$~nm (680~nm) on YIG (SiO$_2$) substrate for the 100~nm thick Cu channel at 150~K in Ref.~\onlinecite{villamor_modulation_2015}, we extract $G_\text{s}=2\times10^{12}$~$\Omega^{-1}\text{m}^{-2}$, which is 5 times larger than their reported $G_\text{r}$ extracted from rotation measurements.
	
	In summary, we have studied the temperature dependence of $G_\text{s}$ and $G_\text{r}$ using the non-local spin valve technique for the Al/YIG interface. From NLSV measurements, we extracted $G_\text{s}$ to be $3.3\times10^{12}$~$\Omega^{-1}\text{m}^{-2}$ at 293~K, which decreases by about $84\%$ at 4.2~K, approximately obeying the $(T/T_\text{c})^{3/2}$ law. While $G_\text{r}$ remains almost constant with the temperature, the value extracted from the modulation of the spin signal ($1\times10^{12}$~$\Omega^{-1}\text{m}^{-2}$) was around 50~times smaller than the calculated value ($5.7\times10^{13}$~$\Omega^{-1}\text{m}^{-2}$). The lower estimate of $G_\text{r}$ from the rotation experiment can be attributed to the formation of an amorphous YIG layer at the interface due to $\text{Ar}^+$ ion milling prior to the evaporation of the Al channel, a consideration missing in the literature so far.

	\begin{acknowledgments}
		We acknowledge the technical support from J.\ G.\ Holstein, H.\ M.\ de Roosz, H.\ Adema, T.\ Schouten and H. de Vries and thank G.\ E.\ W.\ Bauer and F.\ Casanova for discussions. We acknowledge the financial support of the Zernike Institute for Advanced Materials and the Future and Emerging Technologies (FET) programme within the Seventh Framework Programme for Research of the European Commission, under FET-Open Grant No.~618083 (CNTQC). This project is also financed by the NWO Spinoza prize awarded to Prof. B. J. van Wees by the NWO.		
	\end{acknowledgments}


\begin{thebibliography}{44}%
		\makeatletter
		\providecommand \@ifxundefined [1]{%
			\@ifx{#1\undefined}
		}%
		\providecommand \@ifnum [1]{%
			\ifnum #1\expandafter \@firstoftwo
			\else \expandafter \@secondoftwo
			\fi
		}%
		\providecommand \@ifx [1]{%
			\ifx #1\expandafter \@firstoftwo
			\else \expandafter \@secondoftwo
			\fi
		}%
		\providecommand \natexlab [1]{#1}%
		\providecommand \enquote  [1]{``#1''}%
		\providecommand \bibnamefont  [1]{#1}%
		\providecommand \bibfnamefont [1]{#1}%
		\providecommand \citenamefont [1]{#1}%
		\providecommand \href@noop [0]{\@secondoftwo}%
		\providecommand \href [0]{\begingroup \@sanitize@url \@href}%
		\providecommand \@href[1]{\@@startlink{#1}\@@href}%
		\providecommand \@@href[1]{\endgroup#1\@@endlink}%
		\providecommand \@sanitize@url [0]{\catcode `\\12\catcode `\$12\catcode
			`\&12\catcode `\#12\catcode `\^12\catcode `\_12\catcode `\%12\relax}%
		\providecommand \@@startlink[1]{}%
		\providecommand \@@endlink[0]{}%
		\providecommand \url  [0]{\begingroup\@sanitize@url \@url }%
		\providecommand \@url [1]{\endgroup\@href {#1}{\urlprefix }}%
		\providecommand \urlprefix  [0]{URL }%
		\providecommand \Eprint [0]{\href }%
		\providecommand \doibase [0]{http://dx.doi.org/}%
		\providecommand \selectlanguage [0]{\@gobble}%
		\providecommand \bibinfo  [0]{\@secondoftwo}%
		\providecommand \bibfield  [0]{\@secondoftwo}%
		\providecommand \translation [1]{[#1]}%
		\providecommand \BibitemOpen [0]{}%
		\providecommand \bibitemStop [0]{}%
		\providecommand \bibitemNoStop [0]{.\EOS\space}%
		\providecommand \EOS [0]{\spacefactor3000\relax}%
		\providecommand \BibitemShut  [1]{\csname bibitem#1\endcsname}%
		\let\auto@bib@innerbib\@empty
		\bibitem [{\citenamefont {Chumak}\ \emph {et~al.}(2015)\citenamefont {Chumak},
			\citenamefont {Vasyuchka}, \citenamefont {Serga},\ and\ \citenamefont
			{Hillebrands}}]{chumak_magnon_2015}%
		\BibitemOpen
		\bibfield  {author} {\bibinfo {author} {\bibfnamefont {A.~V.}\ \bibnamefont
				{Chumak}}, \bibinfo {author} {\bibfnamefont {V.~I.}\ \bibnamefont
				{Vasyuchka}}, \bibinfo {author} {\bibfnamefont {A.~A.}\ \bibnamefont
				{Serga}}, \ and\ \bibinfo {author} {\bibfnamefont {B.}~\bibnamefont
				{Hillebrands}},\ }\href {\doibase 10.1038/nphys3347} {\bibfield  {journal}
			{\bibinfo  {journal} {Nature Physics}\ }\textbf {\bibinfo {volume} {11}},\
			\bibinfo {pages} {453} (\bibinfo {year} {2015})}\BibitemShut {NoStop}%
		\bibitem [{\citenamefont {Jungwirth}\ \emph {et~al.}(2016)\citenamefont
			{Jungwirth}, \citenamefont {Marti}, \citenamefont {Wadley},\ and\
			\citenamefont {Wunderlich}}]{jungwirth_antiferromagnetic_2016}%
		\BibitemOpen
		\bibfield  {author} {\bibinfo {author} {\bibfnamefont {T.}~\bibnamefont
				{Jungwirth}}, \bibinfo {author} {\bibfnamefont {X.}~\bibnamefont {Marti}},
			\bibinfo {author} {\bibfnamefont {P.}~\bibnamefont {Wadley}}, \ and\ \bibinfo
			{author} {\bibfnamefont {J.}~\bibnamefont {Wunderlich}},\ }\href {\doibase
			10.1038/nnano.2016.18} {\bibfield  {journal} {\bibinfo  {journal} {Nature
					Nanotechnology}\ }\textbf {\bibinfo {volume} {11}},\ \bibinfo {pages} {231}
			(\bibinfo {year} {2016})}\BibitemShut {NoStop}%
		\bibitem [{\citenamefont {Baltz}\ \emph {et~al.}(2018)\citenamefont {Baltz},
			\citenamefont {Manchon}, \citenamefont {Tsoi}, \citenamefont {Moriyama},
			\citenamefont {Ono},\ and\ \citenamefont
			{Tserkovnyak}}]{baltz_antiferromagnetic_2018}%
		\BibitemOpen
		\bibfield  {author} {\bibinfo {author} {\bibfnamefont {V.}~\bibnamefont
				{Baltz}}, \bibinfo {author} {\bibfnamefont {A.}~\bibnamefont {Manchon}},
			\bibinfo {author} {\bibfnamefont {M.}~\bibnamefont {Tsoi}}, \bibinfo {author}
			{\bibfnamefont {T.}~\bibnamefont {Moriyama}}, \bibinfo {author}
			{\bibfnamefont {T.}~\bibnamefont {Ono}}, \ and\ \bibinfo {author}
			{\bibfnamefont {Y.}~\bibnamefont {Tserkovnyak}},\ }\href {\doibase
			10.1103/RevModPhys.90.015005} {\bibfield  {journal} {\bibinfo  {journal}
				{Rev. Mod. Phys.}\ }\textbf {\bibinfo {volume} {90}},\ \bibinfo {pages}
			{015005} (\bibinfo {year} {2018})}\BibitemShut {NoStop}%
		\bibitem [{\citenamefont {Brataas}\ \emph {et~al.}(2000)\citenamefont
			{Brataas}, \citenamefont {Nazarov},\ and\ \citenamefont
			{Bauer}}]{brataas_finite-element_2000}%
		\BibitemOpen
		\bibfield  {author} {\bibinfo {author} {\bibfnamefont {A.}~\bibnamefont
				{Brataas}}, \bibinfo {author} {\bibfnamefont {Y.~V.}\ \bibnamefont
				{Nazarov}}, \ and\ \bibinfo {author} {\bibfnamefont {G.~E.~W.}\ \bibnamefont
				{Bauer}},\ }\href {\doibase 10.1103/PhysRevLett.84.2481} {\bibfield
			{journal} {\bibinfo  {journal} {Phys. Rev. Lett.}\ }\textbf {\bibinfo
				{volume} {84}},\ \bibinfo {pages} {2481} (\bibinfo {year}
			{2000})}\BibitemShut {NoStop}%
		\bibitem [{\citenamefont {Brataas}\ \emph {et~al.}(2006)\citenamefont
			{Brataas}, \citenamefont {Bauer},\ and\ \citenamefont
			{Kelly}}]{brataas_non-collinear_2006}%
		\BibitemOpen
		\bibfield  {author} {\bibinfo {author} {\bibfnamefont {A.}~\bibnamefont
				{Brataas}}, \bibinfo {author} {\bibfnamefont {G.~E.~W.}\ \bibnamefont
				{Bauer}}, \ and\ \bibinfo {author} {\bibfnamefont {P.~J.}\ \bibnamefont
				{Kelly}},\ }\href {\doibase 10.1016/j.physrep.2006.01.001} {\bibfield
			{journal} {\bibinfo  {journal} {Physics Reports}\ }\textbf {\bibinfo {volume}
				{427}},\ \bibinfo {pages} {157} (\bibinfo {year} {2006})}\BibitemShut
		{NoStop}%
		\bibitem [{\citenamefont {Takahashi}\ \emph {et~al.}(2010)\citenamefont
			{Takahashi}, \citenamefont {Saitoh},\ and\ \citenamefont
			{Maekawa}}]{takahashi_spin_2010}%
		\BibitemOpen
		\bibfield  {author} {\bibinfo {author} {\bibfnamefont {S.}~\bibnamefont
				{Takahashi}}, \bibinfo {author} {\bibfnamefont {E.}~\bibnamefont {Saitoh}}, \
			and\ \bibinfo {author} {\bibfnamefont {S.}~\bibnamefont {Maekawa}},\ }\href
		{\doibase 10.1088/1742-6596/200/6/062030} {\bibfield  {journal} {\bibinfo
				{journal} {J. Phys.: Conf. Ser.}\ }\textbf {\bibinfo {volume} {200}},\
			\bibinfo {pages} {062030} (\bibinfo {year} {2010})}\BibitemShut {NoStop}%
		\bibitem [{\citenamefont {Jia}\ \emph {et~al.}(2011)\citenamefont {Jia},
			\citenamefont {Liu}, \citenamefont {Xia},\ and\ \citenamefont
			{Bauer}}]{jia_spin_2011}%
		\BibitemOpen
		\bibfield  {author} {\bibinfo {author} {\bibfnamefont {X.}~\bibnamefont
				{Jia}}, \bibinfo {author} {\bibfnamefont {K.}~\bibnamefont {Liu}}, \bibinfo
			{author} {\bibfnamefont {K.}~\bibnamefont {Xia}}, \ and\ \bibinfo {author}
			{\bibfnamefont {G.~E.~W.}\ \bibnamefont {Bauer}},\ }\href {\doibase
			10.1209/0295-5075/96/17005} {\bibfield  {journal} {\bibinfo  {journal} {EPL}\
			}\textbf {\bibinfo {volume} {96}},\ \bibinfo {pages} {17005} (\bibinfo {year}
			{2011})}\BibitemShut {NoStop}%
		\bibitem [{\citenamefont {Stiles}\ and\ \citenamefont
			{Zangwill}(2002)}]{stiles_anatomy_2002}%
		\BibitemOpen
		\bibfield  {author} {\bibinfo {author} {\bibfnamefont {M.~D.}\ \bibnamefont
				{Stiles}}\ and\ \bibinfo {author} {\bibfnamefont {A.}~\bibnamefont
				{Zangwill}},\ }\href {\doibase 10.1103/PhysRevB.66.014407} {\bibfield
			{journal} {\bibinfo  {journal} {Phys. Rev. B}\ }\textbf {\bibinfo {volume}
				{66}},\ \bibinfo {pages} {014407} (\bibinfo {year} {2002})}\BibitemShut
		{NoStop}%
		\bibitem [{\citenamefont {Ralph}\ and\ \citenamefont
			{Stiles}(2008)}]{ralph_spin_2008}%
		\BibitemOpen
		\bibfield  {author} {\bibinfo {author} {\bibfnamefont {D.~C.}\ \bibnamefont
				{Ralph}}\ and\ \bibinfo {author} {\bibfnamefont {M.~D.}\ \bibnamefont
				{Stiles}},\ }\href {\doibase 10.1016/j.jmmm.2007.12.019} {\bibfield
			{journal} {\bibinfo  {journal} {Journal of Magnetism and Magnetic Materials}\
			}\textbf {\bibinfo {volume} {320}},\ \bibinfo {pages} {1190} (\bibinfo {year}
			{2008})}\BibitemShut {NoStop}%
		\bibitem [{\citenamefont {Brataas}\ \emph
			{et~al.}(2012{\natexlab{a}})\citenamefont {Brataas}, \citenamefont {Kent},\
			and\ \citenamefont {Ohno}}]{brataas_current-induced_2012}%
		\BibitemOpen
		\bibfield  {author} {\bibinfo {author} {\bibfnamefont {A.}~\bibnamefont
				{Brataas}}, \bibinfo {author} {\bibfnamefont {A.~D.}\ \bibnamefont {Kent}}, \
			and\ \bibinfo {author} {\bibfnamefont {H.}~\bibnamefont {Ohno}},\ }\href
		{\doibase 10.1038/nmat3311} {\bibfield  {journal} {\bibinfo  {journal}
				{Nature Materials}\ }\textbf {\bibinfo {volume} {11}},\ \bibinfo {pages}
			{372} (\bibinfo {year} {2012}{\natexlab{a}})}\BibitemShut {NoStop}%
		\bibitem [{\citenamefont {Tserkovnyak}\ \emph {et~al.}(2002)\citenamefont
			{Tserkovnyak}, \citenamefont {Brataas},\ and\ \citenamefont
			{Bauer}}]{tserkovnyak_enhanced_2002}%
		\BibitemOpen
		\bibfield  {author} {\bibinfo {author} {\bibfnamefont {Y.}~\bibnamefont
				{Tserkovnyak}}, \bibinfo {author} {\bibfnamefont {A.}~\bibnamefont
				{Brataas}}, \ and\ \bibinfo {author} {\bibfnamefont {G.~E.~W.}\ \bibnamefont
				{Bauer}},\ }\href {\doibase 10.1103/PhysRevLett.88.117601} {\bibfield
			{journal} {\bibinfo  {journal} {Phys. Rev. Lett.}\ }\textbf {\bibinfo
				{volume} {88}},\ \bibinfo {pages} {117601} (\bibinfo {year}
			{2002})}\BibitemShut {NoStop}%
		\bibitem [{\citenamefont {Deorani}\ and\ \citenamefont
			{Yang}(2013)}]{deorani_role_2013}%
		\BibitemOpen
		\bibfield  {author} {\bibinfo {author} {\bibfnamefont {P.}~\bibnamefont
				{Deorani}}\ and\ \bibinfo {author} {\bibfnamefont {H.}~\bibnamefont {Yang}},\
		}\href {\doibase 10.1063/1.4839475} {\bibfield  {journal} {\bibinfo
				{journal} {Appl. Phys. Lett.}\ }\textbf {\bibinfo {volume} {103}},\ \bibinfo
			{pages} {232408} (\bibinfo {year} {2013})}\BibitemShut {NoStop}%
		\bibitem [{\citenamefont {Nakayama}\ \emph {et~al.}(2013)\citenamefont
			{Nakayama}, \citenamefont {Althammer}, \citenamefont {Chen}, \citenamefont
			{Uchida}, \citenamefont {Kajiwara}, \citenamefont {Kikuchi}, \citenamefont
			{Ohtani}, \citenamefont {Geprägs}, \citenamefont {Opel}, \citenamefont
			{Takahashi}, \citenamefont {Gross}, \citenamefont {Bauer}, \citenamefont
			{Goennenwein},\ and\ \citenamefont {Saitoh}}]{nakayama_spin_2013}%
		\BibitemOpen
		\bibfield  {author} {\bibinfo {author} {\bibfnamefont {H.}~\bibnamefont
				{Nakayama}}, \bibinfo {author} {\bibfnamefont {M.}~\bibnamefont {Althammer}},
			\bibinfo {author} {\bibfnamefont {Y.-T.}\ \bibnamefont {Chen}}, \bibinfo
			{author} {\bibfnamefont {K.}~\bibnamefont {Uchida}}, \bibinfo {author}
			{\bibfnamefont {Y.}~\bibnamefont {Kajiwara}}, \bibinfo {author}
			{\bibfnamefont {D.}~\bibnamefont {Kikuchi}}, \bibinfo {author} {\bibfnamefont
				{T.}~\bibnamefont {Ohtani}}, \bibinfo {author} {\bibfnamefont
				{S.}~\bibnamefont {Geprägs}}, \bibinfo {author} {\bibfnamefont
				{M.}~\bibnamefont {Opel}}, \bibinfo {author} {\bibfnamefont {S.}~\bibnamefont
				{Takahashi}}, \bibinfo {author} {\bibfnamefont {R.}~\bibnamefont {Gross}},
			\bibinfo {author} {\bibfnamefont {G.~E.~W.}\ \bibnamefont {Bauer}}, \bibinfo
			{author} {\bibfnamefont {S.~T.~B.}\ \bibnamefont {Goennenwein}}, \ and\
			\bibinfo {author} {\bibfnamefont {E.}~\bibnamefont {Saitoh}},\ }\href
		{\doibase 10.1103/PhysRevLett.110.206601} {\bibfield  {journal} {\bibinfo
				{journal} {Phys. Rev. Lett.}\ }\textbf {\bibinfo {volume} {110}},\ \bibinfo
			{pages} {206601} (\bibinfo {year} {2013})}\BibitemShut {NoStop}%
		\bibitem [{\citenamefont {Vlietstra}\ \emph
			{et~al.}(2013{\natexlab{a}})\citenamefont {Vlietstra}, \citenamefont {Shan},
			\citenamefont {Castel}, \citenamefont {van Wees},\ and\ \citenamefont
			{Ben~Youssef}}]{vlietstra_spin-hall_2013}%
		\BibitemOpen
		\bibfield  {author} {\bibinfo {author} {\bibfnamefont {N.}~\bibnamefont
				{Vlietstra}}, \bibinfo {author} {\bibfnamefont {J.}~\bibnamefont {Shan}},
			\bibinfo {author} {\bibfnamefont {V.}~\bibnamefont {Castel}}, \bibinfo
			{author} {\bibfnamefont {B.~J.}\ \bibnamefont {van Wees}}, \ and\ \bibinfo
			{author} {\bibfnamefont {J.}~\bibnamefont {Ben~Youssef}},\ }\href {\doibase
			10.1103/PhysRevB.87.184421} {\bibfield  {journal} {\bibinfo  {journal} {Phys.
					Rev. B}\ }\textbf {\bibinfo {volume} {87}},\ \bibinfo {pages} {184421}
			(\bibinfo {year} {2013}{\natexlab{a}})}\BibitemShut {NoStop}%
		\bibitem [{\citenamefont {Uchida}\ \emph {et~al.}(2010)\citenamefont {Uchida},
			\citenamefont {Xiao}, \citenamefont {Adachi}, \citenamefont {Ohe},
			\citenamefont {Takahashi}, \citenamefont {Ieda}, \citenamefont {Ota},
			\citenamefont {Kajiwara}, \citenamefont {Umezawa}, \citenamefont {Kawai},
			\citenamefont {Bauer}, \citenamefont {Maekawa},\ and\ \citenamefont
			{Saitoh}}]{uchida_spin_2010}%
		\BibitemOpen
		\bibfield  {author} {\bibinfo {author} {\bibfnamefont {K.}~\bibnamefont
				{Uchida}}, \bibinfo {author} {\bibfnamefont {J.}~\bibnamefont {Xiao}},
			\bibinfo {author} {\bibfnamefont {H.}~\bibnamefont {Adachi}}, \bibinfo
			{author} {\bibfnamefont {J.}~\bibnamefont {Ohe}}, \bibinfo {author}
			{\bibfnamefont {S.}~\bibnamefont {Takahashi}}, \bibinfo {author}
			{\bibfnamefont {J.}~\bibnamefont {Ieda}}, \bibinfo {author} {\bibfnamefont
				{T.}~\bibnamefont {Ota}}, \bibinfo {author} {\bibfnamefont {Y.}~\bibnamefont
				{Kajiwara}}, \bibinfo {author} {\bibfnamefont {H.}~\bibnamefont {Umezawa}},
			\bibinfo {author} {\bibfnamefont {H.}~\bibnamefont {Kawai}}, \bibinfo
			{author} {\bibfnamefont {G.~E.~W.}\ \bibnamefont {Bauer}}, \bibinfo {author}
			{\bibfnamefont {S.}~\bibnamefont {Maekawa}}, \ and\ \bibinfo {author}
			{\bibfnamefont {E.}~\bibnamefont {Saitoh}},\ }\href {\doibase
			10.1038/nmat2856} {\bibfield  {journal} {\bibinfo  {journal} {Nature
					Materials}\ }\textbf {\bibinfo {volume} {9}},\ \bibinfo {pages} {894}
			(\bibinfo {year} {2010})}\BibitemShut {NoStop}%
		\bibitem [{\citenamefont {Flipse}\ \emph {et~al.}(2014)\citenamefont {Flipse},
			\citenamefont {Dejene}, \citenamefont {Wagenaar}, \citenamefont {Bauer},
			\citenamefont {Youssef},\ and\ \citenamefont {van
				Wees}}]{flipse_observation_2014}%
		\BibitemOpen
		\bibfield  {author} {\bibinfo {author} {\bibfnamefont {J.}~\bibnamefont
				{Flipse}}, \bibinfo {author} {\bibfnamefont {F.}~\bibnamefont {Dejene}},
			\bibinfo {author} {\bibfnamefont {D.}~\bibnamefont {Wagenaar}}, \bibinfo
			{author} {\bibfnamefont {G.}~\bibnamefont {Bauer}}, \bibinfo {author}
			{\bibfnamefont {J.~B.}\ \bibnamefont {Youssef}}, \ and\ \bibinfo {author}
			{\bibfnamefont {B.}~\bibnamefont {van Wees}},\ }\href {\doibase
			10.1103/PhysRevLett.113.027601} {\bibfield  {journal} {\bibinfo  {journal}
				{Phys. Rev. Lett.}\ }\textbf {\bibinfo {volume} {113}},\ \bibinfo {pages}
			{027601} (\bibinfo {year} {2014})}\BibitemShut {NoStop}%
		\bibitem [{\citenamefont {Dejene}\ \emph {et~al.}(2015)\citenamefont {Dejene},
			\citenamefont {Vlietstra}, \citenamefont {Luc}, \citenamefont {Waintal},
			\citenamefont {Ben~Youssef},\ and\ \citenamefont {van
				Wees}}]{dejene_control_2015}%
		\BibitemOpen
		\bibfield  {author} {\bibinfo {author} {\bibfnamefont {F.~K.}\ \bibnamefont
				{Dejene}}, \bibinfo {author} {\bibfnamefont {N.}~\bibnamefont {Vlietstra}},
			\bibinfo {author} {\bibfnamefont {D.}~\bibnamefont {Luc}}, \bibinfo {author}
			{\bibfnamefont {X.}~\bibnamefont {Waintal}}, \bibinfo {author} {\bibfnamefont
				{J.}~\bibnamefont {Ben~Youssef}}, \ and\ \bibinfo {author} {\bibfnamefont
				{B.~J.}\ \bibnamefont {van Wees}},\ }\href {\doibase
			10.1103/PhysRevB.91.100404} {\bibfield  {journal} {\bibinfo  {journal} {Phys.
					Rev. B}\ }\textbf {\bibinfo {volume} {91}},\ \bibinfo {pages} {100404}
			(\bibinfo {year} {2015})}\BibitemShut {NoStop}%
		\bibitem [{\citenamefont {Cornelissen}\ \emph {et~al.}(2015)\citenamefont
			{Cornelissen}, \citenamefont {Liu}, \citenamefont {Duine}, \citenamefont
			{Youssef},\ and\ \citenamefont {Wees}}]{cornelissen_long-distance_2015}%
		\BibitemOpen
		\bibfield  {author} {\bibinfo {author} {\bibfnamefont {L.~J.}\ \bibnamefont
				{Cornelissen}}, \bibinfo {author} {\bibfnamefont {J.}~\bibnamefont {Liu}},
			\bibinfo {author} {\bibfnamefont {R.~A.}\ \bibnamefont {Duine}}, \bibinfo
			{author} {\bibfnamefont {J.~B.}\ \bibnamefont {Youssef}}, \ and\ \bibinfo
			{author} {\bibfnamefont {B.~J.~v.}\ \bibnamefont {Wees}},\ }\href {\doibase
			10.1038/nphys3465} {\bibfield  {journal} {\bibinfo  {journal} {Nature
					Physics}\ }\textbf {\bibinfo {volume} {11}},\ \bibinfo {pages} {1022}
			(\bibinfo {year} {2015})}\BibitemShut {NoStop}%
		\bibitem [{\citenamefont {Cornelissen}\ \emph {et~al.}(2016)\citenamefont
			{Cornelissen}, \citenamefont {Peters}, \citenamefont {Bauer}, \citenamefont
			{Duine},\ and\ \citenamefont {van Wees}}]{cornelissen_magnon_2016}%
		\BibitemOpen
		\bibfield  {author} {\bibinfo {author} {\bibfnamefont {L.~J.}\ \bibnamefont
				{Cornelissen}}, \bibinfo {author} {\bibfnamefont {K.~J.~H.}\ \bibnamefont
				{Peters}}, \bibinfo {author} {\bibfnamefont {G.~E.~W.}\ \bibnamefont
				{Bauer}}, \bibinfo {author} {\bibfnamefont {R.~A.}\ \bibnamefont {Duine}}, \
			and\ \bibinfo {author} {\bibfnamefont {B.~J.}\ \bibnamefont {van Wees}},\
		}\href {\doibase 10.1103/PhysRevB.94.014412} {\bibfield  {journal} {\bibinfo
				{journal} {Phys. Rev. B}\ }\textbf {\bibinfo {volume} {94}},\ \bibinfo
			{pages} {014412} (\bibinfo {year} {2016})}\BibitemShut {NoStop}%
		\bibitem [{\citenamefont {Kalinikos}\ and\ \citenamefont
			{Slavin}(1986)}]{kalinikos_theory_1986}%
		\BibitemOpen
		\bibfield  {author} {\bibinfo {author} {\bibfnamefont {B.~A.}\ \bibnamefont
				{Kalinikos}}\ and\ \bibinfo {author} {\bibfnamefont {A.~N.}\ \bibnamefont
				{Slavin}},\ }\href {\doibase 10.1088/0022-3719/19/35/014} {\bibfield
			{journal} {\bibinfo  {journal} {J. Phys. C: Solid State Phys.}\ }\textbf
			{\bibinfo {volume} {19}},\ \bibinfo {pages} {7013} (\bibinfo {year}
			{1986})}\BibitemShut {NoStop}%
		\bibitem [{\citenamefont {Brataas}\ \emph
			{et~al.}(2012{\natexlab{b}})\citenamefont {Brataas}, \citenamefont
			{Tserkovnyak}, \citenamefont {Bauer},\ and\ \citenamefont
			{Kelly}}]{brataas_spin_2012}%
		\BibitemOpen
		\bibfield  {author} {\bibinfo {author} {\bibfnamefont {A.}~\bibnamefont
				{Brataas}}, \bibinfo {author} {\bibfnamefont {Y.}~\bibnamefont
				{Tserkovnyak}}, \bibinfo {author} {\bibfnamefont {G.~E.~W.}\ \bibnamefont
				{Bauer}}, \ and\ \bibinfo {author} {\bibfnamefont {P.~J.}\ \bibnamefont
				{Kelly}},\ }in\ \href@noop {} {\emph {\bibinfo {booktitle} {Spin
					{Current}}}},\ \bibinfo {editor} {edited by\ \bibinfo {editor} {\bibfnamefont
				{S.}~\bibnamefont {Maekawa}}, \bibinfo {editor} {\bibfnamefont {S.~O.}\
				\bibnamefont {Valenzuela}}, \bibinfo {editor} {\bibfnamefont
				{E.}~\bibnamefont {Saitoh}}, \ and\ \bibinfo {editor} {\bibfnamefont
				{T.}~\bibnamefont {Kimura}}}\ (\bibinfo  {publisher} {Oxford University
			Press},\ \bibinfo {address} {United Kingdom},\ \bibinfo {year} {2012})\ pp.\
		\bibinfo {pages} {87--135}\BibitemShut {NoStop}%
		\bibitem [{\citenamefont {Chen}\ \emph {et~al.}(2013)\citenamefont {Chen},
			\citenamefont {Takahashi}, \citenamefont {Nakayama}, \citenamefont
			{Althammer}, \citenamefont {Goennenwein}, \citenamefont {Saitoh},\ and\
			\citenamefont {Bauer}}]{chen_theory_2013}%
		\BibitemOpen
		\bibfield  {author} {\bibinfo {author} {\bibfnamefont {Y.-T.}\ \bibnamefont
				{Chen}}, \bibinfo {author} {\bibfnamefont {S.}~\bibnamefont {Takahashi}},
			\bibinfo {author} {\bibfnamefont {H.}~\bibnamefont {Nakayama}}, \bibinfo
			{author} {\bibfnamefont {M.}~\bibnamefont {Althammer}}, \bibinfo {author}
			{\bibfnamefont {S.~T.~B.}\ \bibnamefont {Goennenwein}}, \bibinfo {author}
			{\bibfnamefont {E.}~\bibnamefont {Saitoh}}, \ and\ \bibinfo {author}
			{\bibfnamefont {G.~E.~W.}\ \bibnamefont {Bauer}},\ }\href {\doibase
			10.1103/PhysRevB.87.144411} {\bibfield  {journal} {\bibinfo  {journal} {Phys.
					Rev. B}\ }\textbf {\bibinfo {volume} {87}},\ \bibinfo {pages} {144411}
			(\bibinfo {year} {2013})}\BibitemShut {NoStop}%
		\bibitem [{\citenamefont {Xia}\ \emph {et~al.}(2002)\citenamefont {Xia},
			\citenamefont {Kelly}, \citenamefont {Bauer}, \citenamefont {Brataas},\ and\
			\citenamefont {Turek}}]{xia_spin_2002}%
		\BibitemOpen
		\bibfield  {author} {\bibinfo {author} {\bibfnamefont {K.}~\bibnamefont
				{Xia}}, \bibinfo {author} {\bibfnamefont {P.~J.}\ \bibnamefont {Kelly}},
			\bibinfo {author} {\bibfnamefont {G.~E.~W.}\ \bibnamefont {Bauer}}, \bibinfo
			{author} {\bibfnamefont {A.}~\bibnamefont {Brataas}}, \ and\ \bibinfo
			{author} {\bibfnamefont {I.}~\bibnamefont {Turek}},\ }\href {\doibase
			10.1103/PhysRevB.65.220401} {\bibfield  {journal} {\bibinfo  {journal} {Phys.
					Rev. B}\ }\textbf {\bibinfo {volume} {65}},\ \bibinfo {pages} {220401}
			(\bibinfo {year} {2002})}\BibitemShut {NoStop}%
		\bibitem [{\citenamefont {Vlietstra}\ \emph
			{et~al.}(2013{\natexlab{b}})\citenamefont {Vlietstra}, \citenamefont {Shan},
			\citenamefont {Castel}, \citenamefont {Ben~Youssef}, \citenamefont {Bauer},\
			and\ \citenamefont {van Wees}}]{vlietstra_exchange_2013}%
		\BibitemOpen
		\bibfield  {author} {\bibinfo {author} {\bibfnamefont {N.}~\bibnamefont
				{Vlietstra}}, \bibinfo {author} {\bibfnamefont {J.}~\bibnamefont {Shan}},
			\bibinfo {author} {\bibfnamefont {V.}~\bibnamefont {Castel}}, \bibinfo
			{author} {\bibfnamefont {J.}~\bibnamefont {Ben~Youssef}}, \bibinfo {author}
			{\bibfnamefont {G.~E.~W.}\ \bibnamefont {Bauer}}, \ and\ \bibinfo {author}
			{\bibfnamefont {B.~J.}\ \bibnamefont {van Wees}},\ }\href {\doibase
			10.1063/1.4813760} {\bibfield  {journal} {\bibinfo  {journal} {Appl. Phys.
					Lett.}\ }\textbf {\bibinfo {volume} {103}},\ \bibinfo {pages} {032401}
			(\bibinfo {year} {2013}{\natexlab{b}})}\BibitemShut {NoStop}%
		\bibitem [{\citenamefont {Meyer}\ \emph {et~al.}(2014)\citenamefont {Meyer},
			\citenamefont {Althammer}, \citenamefont {Geprägs}, \citenamefont {Opel},
			\citenamefont {Gross},\ and\ \citenamefont
			{Goennenwein}}]{meyer_temperature_2014}%
		\BibitemOpen
		\bibfield  {author} {\bibinfo {author} {\bibfnamefont {S.}~\bibnamefont
				{Meyer}}, \bibinfo {author} {\bibfnamefont {M.}~\bibnamefont {Althammer}},
			\bibinfo {author} {\bibfnamefont {S.}~\bibnamefont {Geprägs}}, \bibinfo
			{author} {\bibfnamefont {M.}~\bibnamefont {Opel}}, \bibinfo {author}
			{\bibfnamefont {R.}~\bibnamefont {Gross}}, \ and\ \bibinfo {author}
			{\bibfnamefont {S.~T.~B.}\ \bibnamefont {Goennenwein}},\ }\href {\doibase
			10.1063/1.4885086} {\bibfield  {journal} {\bibinfo  {journal} {Appl. Phys.
					Lett.}\ }\textbf {\bibinfo {volume} {104}},\ \bibinfo {pages} {242411}
			(\bibinfo {year} {2014})}\BibitemShut {NoStop}%
		\bibitem [{\citenamefont {Villamor}\ \emph {et~al.}(2015)\citenamefont
			{Villamor}, \citenamefont {Isasa}, \citenamefont {Vélez}, \citenamefont
			{Bedoya-Pinto}, \citenamefont {Vavassori}, \citenamefont {Hueso},
			\citenamefont {Bergeret},\ and\ \citenamefont
			{Casanova}}]{villamor_modulation_2015}%
		\BibitemOpen
		\bibfield  {author} {\bibinfo {author} {\bibfnamefont {E.}~\bibnamefont
				{Villamor}}, \bibinfo {author} {\bibfnamefont {M.}~\bibnamefont {Isasa}},
			\bibinfo {author} {\bibfnamefont {S.}~\bibnamefont {Vélez}}, \bibinfo
			{author} {\bibfnamefont {A.}~\bibnamefont {Bedoya-Pinto}}, \bibinfo {author}
			{\bibfnamefont {P.}~\bibnamefont {Vavassori}}, \bibinfo {author}
			{\bibfnamefont {L.~E.}\ \bibnamefont {Hueso}}, \bibinfo {author}
			{\bibfnamefont {F.~S.}\ \bibnamefont {Bergeret}}, \ and\ \bibinfo {author}
			{\bibfnamefont {F.}~\bibnamefont {Casanova}},\ }\href {\doibase
			10.1103/PhysRevB.91.020403} {\bibfield  {journal} {\bibinfo  {journal} {Phys.
					Rev. B}\ }\textbf {\bibinfo {volume} {91}},\ \bibinfo {pages} {020403}
			(\bibinfo {year} {2015})}\BibitemShut {NoStop}%
		\bibitem [{\citenamefont {Muduli}\ \emph {et~al.}(2018)\citenamefont {Muduli},
			\citenamefont {Kimata}, \citenamefont {Omori}, \citenamefont {Wakamura},
			\citenamefont {Dash},\ and\ \citenamefont {Otani}}]{muduli_detection_2018}%
		\BibitemOpen
		\bibfield  {author} {\bibinfo {author} {\bibfnamefont {P.~K.}\ \bibnamefont
				{Muduli}}, \bibinfo {author} {\bibfnamefont {M.}~\bibnamefont {Kimata}},
			\bibinfo {author} {\bibfnamefont {Y.}~\bibnamefont {Omori}}, \bibinfo
			{author} {\bibfnamefont {T.}~\bibnamefont {Wakamura}}, \bibinfo {author}
			{\bibfnamefont {S.~P.}\ \bibnamefont {Dash}}, \ and\ \bibinfo {author}
			{\bibfnamefont {Y.}~\bibnamefont {Otani}},\ }\href {\doibase
			10.1103/PhysRevB.98.024416} {\bibfield  {journal} {\bibinfo  {journal} {Phys.
					Rev. B}\ }\textbf {\bibinfo {volume} {98}},\ \bibinfo {pages} {024416}
			(\bibinfo {year} {2018})}\BibitemShut {NoStop}%
		\bibitem [{\citenamefont {Huang}\ \emph {et~al.}(2012)\citenamefont {Huang},
			\citenamefont {Fan}, \citenamefont {Qu}, \citenamefont {Chen}, \citenamefont
			{Wang}, \citenamefont {Wu}, \citenamefont {Chen}, \citenamefont {Xiao},\ and\
			\citenamefont {Chien}}]{huang_transport_2012}%
		\BibitemOpen
		\bibfield  {author} {\bibinfo {author} {\bibfnamefont {S.~Y.}\ \bibnamefont
				{Huang}}, \bibinfo {author} {\bibfnamefont {X.}~\bibnamefont {Fan}}, \bibinfo
			{author} {\bibfnamefont {D.}~\bibnamefont {Qu}}, \bibinfo {author}
			{\bibfnamefont {Y.~P.}\ \bibnamefont {Chen}}, \bibinfo {author}
			{\bibfnamefont {W.~G.}\ \bibnamefont {Wang}}, \bibinfo {author}
			{\bibfnamefont {J.}~\bibnamefont {Wu}}, \bibinfo {author} {\bibfnamefont
				{T.~Y.}\ \bibnamefont {Chen}}, \bibinfo {author} {\bibfnamefont {J.~Q.}\
				\bibnamefont {Xiao}}, \ and\ \bibinfo {author} {\bibfnamefont {C.~L.}\
				\bibnamefont {Chien}},\ }\href {\doibase 10.1103/PhysRevLett.109.107204}
		{\bibfield  {journal} {\bibinfo  {journal} {Phys. Rev. Lett.}\ }\textbf
			{\bibinfo {volume} {109}},\ \bibinfo {pages} {107204} (\bibinfo {year}
			{2012})}\BibitemShut {NoStop}%
		\bibitem [{\citenamefont {Lu}\ \emph {et~al.}(2013)\citenamefont {Lu},
			\citenamefont {Cai}, \citenamefont {Huang}, \citenamefont {Qu}, \citenamefont
			{Miao},\ and\ \citenamefont {Chien}}]{lu_hybrid_2013}%
		\BibitemOpen
		\bibfield  {author} {\bibinfo {author} {\bibfnamefont {Y.~M.}\ \bibnamefont
				{Lu}}, \bibinfo {author} {\bibfnamefont {J.~W.}\ \bibnamefont {Cai}},
			\bibinfo {author} {\bibfnamefont {S.~Y.}\ \bibnamefont {Huang}}, \bibinfo
			{author} {\bibfnamefont {D.}~\bibnamefont {Qu}}, \bibinfo {author}
			{\bibfnamefont {B.~F.}\ \bibnamefont {Miao}}, \ and\ \bibinfo {author}
			{\bibfnamefont {C.~L.}\ \bibnamefont {Chien}},\ }\href {\doibase
			10.1103/PhysRevB.87.220409} {\bibfield  {journal} {\bibinfo  {journal} {Phys.
					Rev. B}\ }\textbf {\bibinfo {volume} {87}},\ \bibinfo {pages} {220409}
			(\bibinfo {year} {2013})}\BibitemShut {NoStop}%
		\bibitem [{\citenamefont {Cahaya}\ \emph {et~al.}(2014)\citenamefont {Cahaya},
			\citenamefont {Tretiakov},\ and\ \citenamefont {Bauer}}]{cahaya_spin_2014}%
		\BibitemOpen
		\bibfield  {author} {\bibinfo {author} {\bibfnamefont {A.~B.}\ \bibnamefont
				{Cahaya}}, \bibinfo {author} {\bibfnamefont {O.~A.}\ \bibnamefont
				{Tretiakov}}, \ and\ \bibinfo {author} {\bibfnamefont {G.~E.~W.}\
				\bibnamefont {Bauer}},\ }\href {\doibase 10.1063/1.4863084} {\bibfield
			{journal} {\bibinfo  {journal} {Appl. Phys. Lett.}\ }\textbf {\bibinfo
				{volume} {104}},\ \bibinfo {pages} {042402} (\bibinfo {year}
			{2014})}\BibitemShut {NoStop}%
		\bibitem [{\citenamefont {Zhang}\ and\ \citenamefont
			{Zhang}(2012)}]{zhang_spin_2012}%
		\BibitemOpen
		\bibfield  {author} {\bibinfo {author} {\bibfnamefont {S.~S.-L.}\
				\bibnamefont {Zhang}}\ and\ \bibinfo {author} {\bibfnamefont
				{S.}~\bibnamefont {Zhang}},\ }\href {\doibase 10.1103/PhysRevB.86.214424}
		{\bibfield  {journal} {\bibinfo  {journal} {Phys. Rev. B}\ }\textbf {\bibinfo
				{volume} {86}},\ \bibinfo {pages} {214424} (\bibinfo {year}
			{2012})}\BibitemShut {NoStop}%
		\bibitem [{\citenamefont {Bender}\ and\ \citenamefont
			{Tserkovnyak}(2015)}]{bender_interfacial_2015}%
		\BibitemOpen
		\bibfield  {author} {\bibinfo {author} {\bibfnamefont {S.~A.}\ \bibnamefont
				{Bender}}\ and\ \bibinfo {author} {\bibfnamefont {Y.}~\bibnamefont
				{Tserkovnyak}},\ }\href {\doibase 10.1103/PhysRevB.91.140402} {\bibfield
			{journal} {\bibinfo  {journal} {Phys. Rev. B}\ }\textbf {\bibinfo {volume}
				{91}},\ \bibinfo {pages} {140402} (\bibinfo {year} {2015})}\BibitemShut
		{NoStop}%
		\bibitem [{\citenamefont {Xiao}\ and\ \citenamefont
			{Bauer}(2015)}]{xiao_transport_2015}%
		\BibitemOpen
		\bibfield  {author} {\bibinfo {author} {\bibfnamefont {J.}~\bibnamefont
				{Xiao}}\ and\ \bibinfo {author} {\bibfnamefont {G.~E.~W.}\ \bibnamefont
				{Bauer}},\ }\href {http://arxiv.org/abs/1508.02486} {\bibfield  {journal}
			{\bibinfo  {journal} {arXiv:1508.02486 [cond-mat]}\ } (\bibinfo {year}
			{2015})},\ \bibinfo {note} {arXiv: 1508.02486}\BibitemShut {NoStop}%
		\bibitem [{\citenamefont {Das}\ \emph {et~al.}(2016)\citenamefont {Das},
			\citenamefont {Dejene}, \citenamefont {van Wees},\ and\ \citenamefont
			{Vera-Marun}}]{das_anisotropic_2016}%
		\BibitemOpen
		\bibfield  {author} {\bibinfo {author} {\bibfnamefont {K.~S.}\ \bibnamefont
				{Das}}, \bibinfo {author} {\bibfnamefont {F.~K.}\ \bibnamefont {Dejene}},
			\bibinfo {author} {\bibfnamefont {B.~J.}\ \bibnamefont {van Wees}}, \ and\
			\bibinfo {author} {\bibfnamefont {I.~J.}\ \bibnamefont {Vera-Marun}},\ }\href
		{\doibase 10.1103/PhysRevB.94.180403} {\bibfield  {journal} {\bibinfo
				{journal} {Phys. Rev. B}\ }\textbf {\bibinfo {volume} {94}},\ \bibinfo
			{pages} {180403} (\bibinfo {year} {2016})}\BibitemShut {NoStop}%
		\bibitem [{\citenamefont {Das}\ \emph {et~al.}(2017)\citenamefont {Das},
			\citenamefont {Schoemaker}, \citenamefont {van Wees},\ and\ \citenamefont
			{Vera-Marun}}]{das_spin_2017}%
		\BibitemOpen
		\bibfield  {author} {\bibinfo {author} {\bibfnamefont {K.~S.}\ \bibnamefont
				{Das}}, \bibinfo {author} {\bibfnamefont {W.~Y.}\ \bibnamefont {Schoemaker}},
			\bibinfo {author} {\bibfnamefont {B.~J.}\ \bibnamefont {van Wees}}, \ and\
			\bibinfo {author} {\bibfnamefont {I.~J.}\ \bibnamefont {Vera-Marun}},\ }\href
		{\doibase 10.1103/PhysRevB.96.220408} {\bibfield  {journal} {\bibinfo
				{journal} {Phys. Rev. B}\ }\textbf {\bibinfo {volume} {96}},\ \bibinfo
			{pages} {220408} (\bibinfo {year} {2017})}\BibitemShut {NoStop}%
		\bibitem [{\citenamefont {Das}\ \emph {et~al.}(2018)\citenamefont {Das},
			\citenamefont {Liu}, \citenamefont {van Wees},\ and\ \citenamefont
			{Vera-Marun}}]{das_efficient_2018}%
		\BibitemOpen
		\bibfield  {author} {\bibinfo {author} {\bibfnamefont {K.~S.}\ \bibnamefont
				{Das}}, \bibinfo {author} {\bibfnamefont {J.}~\bibnamefont {Liu}}, \bibinfo
			{author} {\bibfnamefont {B.~J.}\ \bibnamefont {van Wees}}, \ and\ \bibinfo
			{author} {\bibfnamefont {I.~J.}\ \bibnamefont {Vera-Marun}},\ }\href
		{\doibase 10.1021/acs.nanolett.8b02114} {\bibfield  {journal} {\bibinfo
				{journal} {Nano Lett.}\ }\textbf {\bibinfo {volume} {18}},\ \bibinfo {pages}
			{5633} (\bibinfo {year} {2018})}\BibitemShut {NoStop}%
		\bibitem [{\citenamefont {Takahashi}\ and\ \citenamefont
			{Maekawa}(2003)}]{takahashi_spin_2003}%
		\BibitemOpen
		\bibfield  {author} {\bibinfo {author} {\bibfnamefont {S.}~\bibnamefont
				{Takahashi}}\ and\ \bibinfo {author} {\bibfnamefont {S.}~\bibnamefont
				{Maekawa}},\ }\href {\doibase 10.1103/PhysRevB.67.052409} {\bibfield
			{journal} {\bibinfo  {journal} {Phys. Rev. B}\ }\textbf {\bibinfo {volume}
				{67}},\ \bibinfo {pages} {052409} (\bibinfo {year} {2003})}\BibitemShut
		{NoStop}%
		\bibitem [{\citenamefont {Žutić}\ \emph {et~al.}(2004)\citenamefont
			{Žutić}, \citenamefont {Fabian},\ and\ \citenamefont
			{Das~Sarma}}]{zutic_spintronics:_2004}%
		\BibitemOpen
		\bibfield  {author} {\bibinfo {author} {\bibfnamefont {I.}~\bibnamefont
				{Žutić}}, \bibinfo {author} {\bibfnamefont {J.}~\bibnamefont {Fabian}}, \
			and\ \bibinfo {author} {\bibfnamefont {S.}~\bibnamefont {Das~Sarma}},\ }\href
		{\doibase 10.1103/RevModPhys.76.323} {\bibfield  {journal} {\bibinfo
				{journal} {Rev. Mod. Phys.}\ }\textbf {\bibinfo {volume} {76}},\ \bibinfo
			{pages} {323} (\bibinfo {year} {2004})}\BibitemShut {NoStop}%
		\bibitem [{\citenamefont {Zholud}\ \emph {et~al.}(2017)\citenamefont {Zholud},
			\citenamefont {Freeman}, \citenamefont {Cao}, \citenamefont {Srivastava},\
			and\ \citenamefont {Urazhdin}}]{zholud_spin_2017}%
		\BibitemOpen
		\bibfield  {author} {\bibinfo {author} {\bibfnamefont {A.}~\bibnamefont
				{Zholud}}, \bibinfo {author} {\bibfnamefont {R.}~\bibnamefont {Freeman}},
			\bibinfo {author} {\bibfnamefont {R.}~\bibnamefont {Cao}}, \bibinfo {author}
			{\bibfnamefont {A.}~\bibnamefont {Srivastava}}, \ and\ \bibinfo {author}
			{\bibfnamefont {S.}~\bibnamefont {Urazhdin}},\ }\href {\doibase
			10.1103/PhysRevLett.119.257201} {\bibfield  {journal} {\bibinfo  {journal}
				{Phys. Rev. Lett.}\ }\textbf {\bibinfo {volume} {119}},\ \bibinfo {pages}
			{257201} (\bibinfo {year} {2017})}\BibitemShut {NoStop}%
		\bibitem [{\citenamefont {Kim}\ \emph {et~al.}(2014)\citenamefont {Kim},
			\citenamefont {Khim}, \citenamefont {Chun}, \citenamefont {Jo}, \citenamefont
			{Balicas}, \citenamefont {Yi}, \citenamefont {Cheong}, \citenamefont
			{Harrison}, \citenamefont {Batista}, \citenamefont {Hoon~Han},\ and\
			\citenamefont {Hoon~Kim}}]{kim_manifestation_2014}%
		\BibitemOpen
		\bibfield  {author} {\bibinfo {author} {\bibfnamefont {J.~W.}\ \bibnamefont
				{Kim}}, \bibinfo {author} {\bibfnamefont {S.}~\bibnamefont {Khim}}, \bibinfo
			{author} {\bibfnamefont {S.~H.}\ \bibnamefont {Chun}}, \bibinfo {author}
			{\bibfnamefont {Y.}~\bibnamefont {Jo}}, \bibinfo {author} {\bibfnamefont
				{L.}~\bibnamefont {Balicas}}, \bibinfo {author} {\bibfnamefont {H.~T.}\
				\bibnamefont {Yi}}, \bibinfo {author} {\bibfnamefont {S.-W.}\ \bibnamefont
				{Cheong}}, \bibinfo {author} {\bibfnamefont {N.}~\bibnamefont {Harrison}},
			\bibinfo {author} {\bibfnamefont {C.~D.}\ \bibnamefont {Batista}}, \bibinfo
			{author} {\bibfnamefont {J.}~\bibnamefont {Hoon~Han}}, \ and\ \bibinfo
			{author} {\bibfnamefont {K.}~\bibnamefont {Hoon~Kim}},\ }\href {\doibase
			10.1038/ncomms5419} {\bibfield  {journal} {\bibinfo  {journal} {Nature
					Communications}\ }\textbf {\bibinfo {volume} {5}},\ \bibinfo {pages} {4419}
			(\bibinfo {year} {2014})}\BibitemShut {NoStop}%
		\bibitem [{\citenamefont {Cherepanov}\ \emph {et~al.}(1993)\citenamefont
			{Cherepanov}, \citenamefont {Kolokolov},\ and\ \citenamefont
			{L'vov}}]{cherepanov_saga_1993}%
		\BibitemOpen
		\bibfield  {author} {\bibinfo {author} {\bibfnamefont {V.}~\bibnamefont
				{Cherepanov}}, \bibinfo {author} {\bibfnamefont {I.}~\bibnamefont
				{Kolokolov}}, \ and\ \bibinfo {author} {\bibfnamefont {V.}~\bibnamefont
				{L'vov}},\ }\href {\doibase 10.1016/0370-1573(93)90107-O} {\bibfield
			{journal} {\bibinfo  {journal} {Physics Reports}\ }\textbf {\bibinfo {volume}
				{229}},\ \bibinfo {pages} {81} (\bibinfo {year} {1993})}\BibitemShut
		{NoStop}%
		\bibitem [{\citenamefont {Vélez}\ \emph {et~al.}(2016)\citenamefont {Vélez},
			\citenamefont {Bedoya-Pinto}, \citenamefont {Yan}, \citenamefont {Hueso},\
			and\ \citenamefont {Casanova}}]{velez_competing_2016}%
		\BibitemOpen
		\bibfield  {author} {\bibinfo {author} {\bibfnamefont {S.}~\bibnamefont
				{Vélez}}, \bibinfo {author} {\bibfnamefont {A.}~\bibnamefont
				{Bedoya-Pinto}}, \bibinfo {author} {\bibfnamefont {W.}~\bibnamefont {Yan}},
			\bibinfo {author} {\bibfnamefont {L.~E.}\ \bibnamefont {Hueso}}, \ and\
			\bibinfo {author} {\bibfnamefont {F.}~\bibnamefont {Casanova}},\ }\href
		{\doibase 10.1103/PhysRevB.94.174405} {\bibfield  {journal} {\bibinfo
				{journal} {Phys. Rev. B}\ }\textbf {\bibinfo {volume} {94}},\ \bibinfo
			{pages} {174405} (\bibinfo {year} {2016})}\BibitemShut {NoStop}%
		\bibitem [{\citenamefont {Casanova}(2018)}]{casanova_private_2018}%
		\BibitemOpen
		\bibfield  {author} {\bibinfo {author} {\bibfnamefont {F.}~\bibnamefont
				{Casanova}},\ }\href@noop {} {\enquote {\bibinfo {title} {Private
					communication},}\ } (\bibinfo {year} {2018})\BibitemShut {NoStop}%
		\bibitem [{\citenamefont {Wesenberg}\ \emph {et~al.}(2017)\citenamefont
			{Wesenberg}, \citenamefont {Liu}, \citenamefont {Balzar}, \citenamefont
			{Wu},\ and\ \citenamefont {Zink}}]{wesenberg_long-distance_2017}%
		\BibitemOpen
		\bibfield  {author} {\bibinfo {author} {\bibfnamefont {D.}~\bibnamefont
				{Wesenberg}}, \bibinfo {author} {\bibfnamefont {T.}~\bibnamefont {Liu}},
			\bibinfo {author} {\bibfnamefont {D.}~\bibnamefont {Balzar}}, \bibinfo
			{author} {\bibfnamefont {M.}~\bibnamefont {Wu}}, \ and\ \bibinfo {author}
			{\bibfnamefont {B.~L.}\ \bibnamefont {Zink}},\ }\href {\doibase
			10.1038/nphys4175} {\bibfield  {journal} {\bibinfo  {journal} {Nature
					Physics}\ }\textbf {\bibinfo {volume} {13}},\ \bibinfo {pages} {987}
			(\bibinfo {year} {2017})}\BibitemShut {NoStop}%
	\end{thebibliography}
	
	%

\end{document}